# Omid Karami

# Mina Mahmoudi

**Estimating Trade-Related Adjustment Costs in the Agricultural Sector in Iran**

## Abstract

Tariff liberalization and its impact on tax revenue is an important consideration for developing countries, because they are increasingly facing the difficult task of implementing and harmonizing regional and international trade commitments. The tariff reform and its costs for Iranian government is one of the issues that are examined in this study. Another goal of this paper is, estimating the cost of trade liberalization. On this regard, imports value of agricultural sector in Iran in 2010 was analyzed according to two scenarios. For reforming nuisance tariff, a VAT policy is used in both scenarios. In this study, TRIST method is used. In the first scenario, imports' value decreased to a level equal to the second scenario and higher tariff revenue will be created. The results show that reducing the average tariff rate does not always result in the loss of tariff revenue. This paper is a witness that different forms of tariff can generate different amount of income when they have same level of liberalization and equal effect on producers. Therefore, using a good tariff regime can help a government to generate income when increases social welfare by liberalization.

1. Introduction

Measuring the benefits of trade reform has been a frustrating endeavor. Although the discussion of trade policy at times gives the impression that a liberal trade regime can do wonders for a country's economy, and most observers believe firmly that trade reform is beneficial, yet systematic attempts at quantification fail to single out trade policy as a major factor in economic growth. The channels through which trade liberalization could bring benefits are broadly these: improved resource allocation in line with social marginal costs and benefits; access to better technologies, inputs and intermediate goods; an economy better able to take advantage of economies of scale and scope; greater domestic competition; availability of favorable growth externalities, like the transfer of know-how; and a shakeup of industry

27

that may create a Schumpeterian environment especially conducive to growth (Dornbusch, 1992).

On the other hand, tax revenue and tax policy are used to obtain national objectives. Normally, governments use tax revenue to provide the goods and services required to enhance economic growth and development, correct market failures, redistribute income and wealth and maintain macroeconomic stability. The latter is particularly important in small open economies which are highly dependent on the world economy such as developing countries (IMF, 2009).

In developed countries, the personal income tax raises a large share of total tax revenue. In developing countries, however, it encounters serious difficulties as they are often characterized by large informal sectors. A large part of the labor force is employed in the agricultural sector or in small informal enterprises where incomes fluctuate and records are not accurately kept (Tanzi and Howell, 2001). The relative size of this informal sector is often three or four times larger in developing countries than in developed countries (Bird et al, 2008). Since ways of effectively taxing the informal transactions has not yet been found, the base for personal income tax is narrow in developing countries and levying high taxes on that base may enhance the existing distortions. The large informal sector, which also complicates the taxation on consumption, is one of the factors restricting the tax policy design of developing countries (Tanzi and Howell, 2001).

Liberalizing trade and reforming the domestic tax system in order to recover the resulting government revenue loss could be a growth-enhancing process if carried out successfully. Trade theory tells us that trade liberalization leads to increased GDP and trade flows, which raises welfare through increased consumption opportunities. If the tax system is efficient enough, tax revenue grows in the same proportion and captures the gains from trade in the state budget. Generally, developing countries are advised to substitute their import duties with indirect taxes. As the tariff elimination leads to lower import prices, provided that there is competition keeping the producer from maintaining a higher price level, a commodity tax can be added without making the consumer worse off (Baunsgaard and Michael, 2005). Moreover, overall domestic consumption is broader than imports, implying that a shift from trade taxes to consumption taxes on substantially all commodities broadens the tax base and enhances horizontal equity in the system.

The need of domestic tax reform in the context of trade liberalization is a result of tariff revenue being reduced. As tariffs are eliminated, revenue has to be collected from other sources. However, when tariffs are not eliminated but reduced, some of the resulting fiscal impact may be offset by an increase in imports (Hallaert, 2008). Even the revenue loss from an elimination of tariffs can be offset by the increase in imports when a VAT is levied on the import good and the increased import flow hence increases VAT revenue. This reasoning stems from the general assumption of trade liberalization leading to increased trade volumes. This assumption is in turn based on the assumption that tariff reductions lead to price reductions, boosting domestic demand for import goods. It should be noted that in order for this to be true, the import goods need to be somewhat price elastic. Moreover, one should keep in mind that lacking competition could hinder the tariff reductions from turning into price reductions; such market failure may allow companies to increase profits instead of lowering prices. In the cases where insufficient competition allows the import price reduction to accrue to the companies in the shape of profits, consumption is left unchanged and hence, so is VAT revenue. However, if the taxation is effective enough, the enterprises' increased profits could contribute to an increase in revenue from the corporate income tax, reducing the revenue loss (Ferreira, et al, 2006).

In particular, tariff liberalization and its impact on tax revenue is an important consideration for developing countries, because they are increasingly facing the difficult task of implementing and harmonizing regional and international trade commitments. For many developing countries, tariffs and other import taxes provide significant sources of revenue. Having the capacity to equip policymakers with the means to project the adjustment impacts of trade reform is thus integral to developing effective trade reform strategies (Hamolton, 2010).

In Iran, as a developing country, because of government intervention in economic activities specifically in the field of trade policies, tariff as a trade instrument has not been applied correctly. Government usually uses non-tariff instruments for controlling imports and dealing with other countries has some rules which included in tariff applying too. Tariff regime in Iran is characterized by high average tariff (nominal and import-weighted tariff), an extremely high dispersion of tariff levels across tariff lines and a significant amount of tariff escalation (Chemingui and Dessus, 2008). Iranian economic history has been influenced by import substitution industrialization approach to develop and consequently agriculture sector, which a large part of the labor force is employed in, has been ignored compared to industry sector. Therefore, knowing the effects of trade adjustments in Iran is important especially in the large agricultural sector.

Many general equilibrium studies have assessed the economic impacts of tariff reform and domestic complementary policies in developing countries. In a static general equilibrium model for Syria, B. Lucke (2001) studies different scenarios of preferential trade liberalization with the EU, and focuses on the effects of tariff reform on government budget. The study finds that government revenue losses caused by reduction in the EU import duties are fairly large, but still manageable. Omar Feraboli (2003) in a study on the effects of Jordan Association Agreement with the EU on macroeconomic variables found that the impact of trade liberalization on welfare is positive under all scenarios and also trade liberalization reduces government revenue, due to foregone import duties. Nashashibi (2002) provided a detailed outlook of revenue performance in Southern Mediterranean Arab countries. He acknowledged that there is higher trade protection in these countries than in other regions and pointed to the expectation that trade liberalization would lead to further decreases in revenues. Tosun (2005) in a study on Middle East and North Africa countries indicated that none of the major tax revenue sources were significantly impacted by the increased trade openness in the post-1986 period. In this paper, evidence is provided to show that there was a statistically significant move to domestic taxes on goods and services in trade liberalizing non-OECD countries. While this finding is supported for non-OECD countries in general and for other non-OECD countries excluding the MENA countries, it failed to materialize for the MENA countries. This suggests that the composition of available tax instruments in the MENA countries did not change in favor of those taxes that are thought to be welfare improving compared with international trade taxes. This could pose several problems for the MENA countries, such as lower economic growth, lower revenue available for economic development and high unemployment due to lack of job opportunities.

In this paper, we want to assess the effects of trade liberalization and changes in tariff regime on the import of agricultural products in Iran and also its impact on tax revenue as a large part of government income.

## 2. Methodology

In this paper we used a methodology that Work Bank has published in 2009. The name of the methodology is Tariff Reform Impact Simulation Tool (TRIST). The trade model in TRIST is based on the standard Armington (1969) assumption of imperfect substitution between imports from different sources. The model does not allow for direct substitution between products. The trade response to a change in tariffs for a given product from a given exporter is calculated based on the resulting percentage change in the duty inclusive price. One of the advantages of TRIST is that it uses actual import transactions data at the tariff line level to project trade reform adjustment costs. As such, it distinguishes between collected tariff revenue (which is calculated based on applied tariffs) and statutory tariff revenue. This allows for projections to be made using data on actual revenue collected as opposed to using revenue data based on statutory rates which do not reflect the tariff exemptions that are applied.

Another feature of TRIST is that specific country groups can be developed to reflect relevant trading blocs and agreements. This enables tailored scenarios to be created for specific country groups while leaving trade with other partners unchanged. The trade model in TRIST is based on five core assumptions: First, the model is derived from standard consumer demand theory and utilizes elasticities to determine the magnitude of the demand response to the price changes that result from a tariff reform. Second, the calculations are based on the standard Armington (1969) assumption of imperfect substitution between imports from different trading partners since consumers distinguish products by the place of production. Third, the model does not allow for direct substitution between different products. Fourth, it is assumed that all changes in tariffs are fully passed on and that the world price remains unchanged. Fifth, the trade model in TRIST is a partial equilibrium model that treats demand for each product in isolation from the rest of the economy (Fernandez and Rodrik, 1997). TRIST is an Excel based tool that predicts the impact of tariff reform scenarios on the basis of a simple partial equilibrium model. It consists of two Excel files: the first is the Data Aggregation Tool which organizes and appropriately formats the data to be imported into the second, the Simulation Tool. It is assumed that all changes in tariffs are fully passed on and that the world price remains unchanged (infinite supply elasticity). The calculation of the price change depends on how a country applies its tariffs, excise and VAT. In most countries, tariffs are collected as a percentage of the C.I.F import value, excise taxes on the tariff inclusive C.I.F import value and VAT on the tariff and excise inclusive C.I.F import value. Thus, for a change in the tariff (with VAT and excise rates unchanged), the percentage change in the price of good i from exporter j is calculated as follows (subscript i is omitted on all arguments in the formula (Brenton et al, 2009):

$$\frac{\Delta p_j}{P_j^{old}} = \frac{[P_j^{new}/P_{wld}] - [P_j^{old}/P_{wld}]}{[P_j^{old}/P_{wld}]} = \frac{(1+t_j^{new})(1+ext_j)(1+vat_j) - (1+t_j^{old})(1+ext_j)(1+vat_j)}{(1+t_j^{old})(1+ext_j)(1+vat_j)}$$

$$= \frac{t_j^{new} - t_j^{old}}{(1+t_j^{old})} \quad (1)$$

Where:
- change in price of imports from country j
- price of imports from j before tariff reform
- price of imports from j after tariff reform
- world market price
- tariff rate applied to imports from country j before reform
- tariff rate applied to imports from country j after reform
- excise tax rate applied to imports from j
- VAT rate applied to imports from j

The trade response for a particular product is modeled in three consecutive steps: (1) the substitution between different exporters following changes in relative prices of different suppliers due to preferential tariff or duty changes, (2) the substitution between imports and domestic output as the relative price of overall imports of the product changes relative to domestically produced goods, and (3) a demand effect whereby consumption of the product changes in response to a change in the overall price of the product[8]. In this survey only first effect is calculated and others are supposed fixed.

In the first stage we model the allocation of given expenditure on imports of a product across different country suppliers and how this allocation changes when tariffs and duties are amended (Brenton et al, 2009):

$$q_j^{ES} = \left[\frac{\Delta P_j}{P_j^{old}} * Y_j^{ES} + 1\right] q_j^{old} * \frac{\Sigma_{j=1,\ldots,n}(q_j^{old})}{\Sigma_{j=1,\ldots,n}\left(\left[\frac{\Delta P_j}{P_j^{old}}*+1\right]q_j^{old}\right)} \qquad (2)$$

Where:
- imported quantity from j after exporter substitution step
- imported quantity from j before reform
- exporter substitution elasticity for imports from country j

### 3. Results and Discussion

Required data for this study includes: the import quantity of products from partners, C.I.F prices of import products, added value of products, substitution elasticity between domestic and import products and substitution elasticity between exporters. We have used imports and tariff statistics data of Iran for the year 2010. Elasticity of substitution between different exporters is supposed to be 1.5.

Two scenarios are assessing in this paper:
1. Fixing tariff rates above 40% at 35%, fixing tariff rates above 20% and below or equal 40% at 30%, raising bands above 10% (5%) and below or equal 20% (10%) to 20% (10%), and fixing all nuisance tariffs at 5%.
2. Eliminating tariff rates above 40%, fixing tariff rates above 20% and below or equal 40% at 35%, fixing bands above 10% (5%) and below or equal 20% (10%) to 20% (10%) and increases all nuisance tariffs to 5%.

**Table1 Ranges of Tariffs in Two Scenarios**

| Existing tariff | First scenario | Second scenario |
|---|---|---|
| Above 40% | 35% | 0% |
| Between 20% & 40% | 30% | 35% |
| Between 10% & 20% | 10% | 20% |
| Between 5% & 10% | 5% | 10% |
| Below 5% (nuisance) | 5% | 5% |

Unlike the papers done in this area, this paper doesn't suggest to eliminate nuisance tariff. We can use an added-value instead of them. Importing of agricultural products in Iran is based on "political considerations" not "economic considerations". So, in this survey another scenario for assessing the effects of a Preferential Trading Arrangements (PTA)is not studied. Now we want to see that between two scenarios which can change imports of agricultural products by the first effect in TRIST.

The value of agricultural import products and the trade partners are presented in table 2. UAE is the country that exports the most agricultural products to Iran. It is clear that these products have not produced in UAE and there is no statistics to understand what the origin countries of these products are. Customs in some countries separate goods that are imported for re-export from that are for domestic consumption. For example, Lim and Saborowski(2010) in a paper on estimating trade related adjustment costs in Syria used CPC code for separating these two kinds of goods, but Iran's customs does not separate these goods. HS code is used in Iran.

**Table 2 Value of Agricultural Products Imports and the Share Of Trade Partners**

| Country | Export to Iran(1000 Billion Rials) | Share (%) |
|---|---|---|
| UAE | 34.9 | 37 |
| Swiss | 11.9 | 13 |
| Nederland | 10.7 | 11 |
| Brazil | 6.1 | 6 |
| Germany | 4.9 | 5 |
| Total | 94.1 | 100 |

Table3 shows tariff income for different range of tariffs. The most income is generated from above 40% tariff rates. Nuisance tariffs are generated 19.8 percent of tariff incomes. This income doesn't cover its costs. So, one of the changes should be done in nuisance tariff, a change from which this income can be generated without any costs.

**Table 3 Tariff Income in Different Ranges of Tariffs**

| Tariff ranges | Income (1000 billion Rials) | Share (%) |
|---|---|---|
| Above 40% | 6.95 | 54.1 |
| Between 20% & 40% | 1.27 | 9.9 |
| Between 10% & 20% | 1.95 | 15.1 |
| Between 5% & 10% | 0.15 | 1.1 |
| Below 5%(nuisance) | 2.54 | 19.8 |
| Total | 12.86 | 100 |

Using TRIST method, changes in imports of agricultural products in different scenarios can be calculated as presented in table 4 and 5.

**Table 4 Results of First Scenario in TRIST** (values are in Billiards Rial)

| Variables | Value |
|---|---|
| Import value (before) | 94.1 |
| Import value (after) | 119 |
| Changes | 24.9 |
| Tariff income (before) | 12.9 |
| Tariff income (after) | 64.8 |
| Changes | 51.9 |
| Average tariff (before) | 24 |
| Average tariff (after) | 16 |
| Changes | -8 |

**Table 5 Results of Second Scenario in TRIST** (values are in Billiards Rial)

| Variables | Value |
|---|---|
| Import value (before)[*] | 94.1 |
| Import value (after) | 119 |
| Changes | 24.9 |
| Tariff income (before) | 12.9 |
| Tariff income (after) | 0.2 |
| Changes | -12.7 |
| Average tariff (before) | 24 |
| Average tariff (after) | 12 |
| Changes | -12 |

As it can be seen in Table 4 and 5, values of agricultural product imports in both scenarios is equal and it is more than before. This shows that both scenarios have equal effects on producers and have lowered average tariff than before. So, both have a degree of liberalization. Also, the first scenario shows an increase in government income while the second one decreases it. This shows that different types of tariff are important for government income. Considering that both scenarios have equal import changes and the first one generates more income for government, the first scenario is better than the second one. Increase in government income when average of tariffs is decreased (according to the first scenario), is a surprising result. It is known that especially in developing countries openness has harmful results on economy and would decrease government income, but openness to international trade helps increasing tax incomes possibly by increasing employment, wage level, and corporate profits. It also has a contribution to the consumption tax possibly by spurring flows of goods within the country. In many studies, positive relationship between the degree of trade openness and trade tax means that openness possibly is a stimulus to higher volume of trade between countries and consequently increases trade tax receipts at the current level of the tariff rate. These are some of the papers that have obtained the same result: Brenton et.al (2007) used a partial equilibrium to estimate revenue impacts of tariff reform in COMESA. Their study seeks to contribute to discussions concerning the potential impacts on tax revenues resulting from a move to a customs union CU and the implementation of an EPA. Results of this study showed that if all tariffs were to be removed on imports from EU, tariff revenue would be increased by 24.6% for Zambia and 26.2% for Malawi. Pupongsak (2009) investigated the trade and revenue impact of trade liberalization and argued that the effect of trade openness on both trade tax and domestic taxes emphasizes the fact that, for low and middle income countries, not only is their trade sector highly dependent on international sector, but also their entire economic structure will be affected if there is any change in countries' international trade system. Also a change which leads to an increase in trade volume will consequently benefit these countries' taxation. Thus, although overall results suggest that trade liberalization via increasing trade openness generally has a contribution to taxation in all countries; the degree of its benefit depends on the country's level of economic development and economic structure. In the lecture review, there are also some studies that have vice versa results. For example, Andriamananjara et.al (2009) assessed the economic impact of an EPA on Nigeria and showed that an EPA will decrease tariff and VAT revenue at least -13.5%. So, effect of liberalization differs from one country to another and also from one scenario to another. Decreasing in tariff amount can decreases government income in a country or increases that. Our results show that for Iranian economy, a decrease in tariff according to the first scenario can increase the government revenue, while by the second one it will decrease the government income.

## 4. Conclusion

Governments in developing countries usually believe that liberalization and decreasing tariffs amounts has a big cost. So, some studies are needed for estimating these costs. Word Bank suggested a method named TRIST for using different scenarios. The trade model in TRIST is based on the standard Armington (1969) assumption of imperfect substitution between imports from different sources. The trade response for a particular product is modeled in three consecutive steps in TRIST: (1) the substitution between different exporters, (2) the substitution between imports and domestic output, and (3) a demand effect whereby consumption of the product changes. In this study, trade related adjustment costs for Iran is estimated. From the three effects that are mentioned, only first one is estimated. Others are supposed fixed. Two scenarios were analyzed for changing tariff amounts. Two scenarios were assessed:

    1. Fixing tariff rates above 40% at 35%, fixing tariff rates above 20% and below or equal 40% at 30%, raising bands above 10% (5%) and below or equal 20% (10%) to 20% (10%), and fixing all nuisance tariffs at 5%.
    2. Eliminating tariff rates above 40%, fixing tariff rates above 20% and below or equal 40% at 35%, fixing bands above 10% (5%) and below or equal 20% (10%) to 20% (10%) and increases all nuisance tariffs to 5%.

Both of them had an equal tariff average.

After using TRIST, results showed that values of imports in both scenarios are equal and they are more than before. This shows that both scenarios have equal effect on producers. First scenario shows an increase in income of government when second one decreases it. So, different forms of tariff are important for generating income for the government. Both scenarios have an equal import changes and the first one generates more income for government, so the first one is better. Both scenarios have lowered average tariff than before. So, both of them have a degree of liberalization. This paper is a witness that different forms of tariff can generate different amount of income when they have same level of liberalization and equal effect on producers. Therefore, using a good tariff regime can help a government to generate income when increases social welfare by liberalization. So, according to the results these suggestions are for the government:

> 1. All kind of liberalizations don't have a decreasing effect on government revenue. This effect should be estimated before implementing every policy.
> 2. Nuisance tariffs don't generate revenue for the government. A policy that eliminates them and uses VAT should be implemented.

As mentioned above, in this paper only one of the effects in TRIST model is estimated. It can be a weakness of this paper. So, estimating other two effects can be estimated by other researchers. It can be a good subject for other papers.

**References**

**Abedini, J. and Péridy, N. J. (2008),** "The Greater Arab Free Trade Area (GAFTA): An estimation of its trade effects," *Journal of Economic Integration* 23(4), pp. 848–872.

**Andriamananjara.S, Brenton.p, von Uexküll.J.E, and Walkenhorst. p. (2009),** *Assessing the Economic Impacts of an Economic Partnership Agreement on Nigeria*, The World Bank Africa Region Africa Technical Families.

**Armington, P. S. (1969),** "A theory of demand for products distinguished by place of production," *IMF Staff Papers*, 16(1), pp. 159–178.

**Atici, C., and Kennedy, P. L. (2005),** "Tradeoffs between income distribution and welfare: The case of Turkey's integration into the European Union," *Journal of Policy Modeling*, 27(5), pp. 553–563.

**Baunsgaard, T. and Michael K. (2005)**, *Tax Revenue and (or?) Trade Liberalization*. IMF Working Paper.

**Bevan, D. L. (1999),** "Trade liberalization and the budget deficit," *Journal of Policy Modeling* 21(6), pp. 653–694.

**Bird, Richard M. and Zolt, Eric M.(2008),** *Technology and Taxation in Developing Countries: From Hand to Mouse*. Mimeo.

**Brenton, P., Saborowski, C., Staritz, C., and von Uexkull, E. (2009),** *Assessing the adjustment costs of trade policy reform using TRIST (Tariff Reform Impact Simulation Tool)*, World Bank Policy Research Paper 5045.Washington, DC: The World Bank.

**Chemingui, M. A., and Dessus, S. C. (2008),** "Assessing non-tariff barriers in Syria," *Journal of Policy Modeling*, 30(5), pp. 917–928.

**Davidson, C., Matusz, S. J., and Nelson, D. R. (2007),** "Can compensation save free trade?," *Journal of International Economics* 71(1), pp. 167–186.

**Dornbusch, R., (1992),** "The Case for Trade Liberalization in Developing Countries," *Journal of Economic Perspectives* 6(1), pp. 69-85.

**Feizabadi.Y, (2007),** *Political tariff protection in Iran's Agriculture Sector*, Paper prepared for presentation at the 106th seminar of the EAAE.

**Fernández, R., and Rodrik, D. (1991),** "Resistance to reform: Status quo bias in the presence of individual-specific uncertainty," *American Economic Review* 81(5), pp. 1146–1155.

**Ferreira, A. Miranda, R., António C., Malta, J. C. (2006),** *Estudo de Impacto de um Acorde de ParceriaEconómica*, SociedadeNacional de Empreendimentos, Portugal.

**Gaitán, B., and Lucke. (2007),** "The Barcelona initiative and the importance of NTBs: A dynamic CGE-analysis for Syria," *International Economics and Economic Policy* 4(1), pp. 33-59.

**Grossman, G. M., and Helpman, E. (1994),** "Protection for sale," *American Economic Review* 84(4), pp. 833–850.

**Hallaert, J-J, (2008),** *How does a domestic tax reform affect protection against imports? The case of the Republic of Madagascar*, IMF Working Paper.

**Hamolton.A, (2010),** *Discussing Existing TRIST Tools for Twelve Developing Countries: Bolivia, Burundi, Ethiopia, Jordan, Kenya, Madagascar, Malawi, Mozambique, Nigeria, Seychelles, Tanzania and Zambia*, World Bank Policy Research Paper 5045.Washington, DC: The World bank.

**Hosoe, N. (2001),** "A general equilibrium analysis of Jordan's trade liberalization," *Journal of Policy Modeling* 23(6), pp. 595–600.

**Ianchovichina, E. (2004),** "Trade policy analysis in the presence of duty drawbacks," *Journal of Policy Modeling* 26(3), pp. 353–371.

**IMF (2009),** *Cape Verde: Seventh Review Under the Policy Support Instrument, approved by SharminiCoorey and Aasim Husain*, IMF Country Report No. 09/328, Washington D.C.

**Kouparitsas, M. A. (2001),** "Should trade barriers be phased-out slowly? A case study of North America," *Journal of Policy Modeling* 23(8), pp. 875–900.

**Lim.J.J and Saborowski.C, (2010),** "Estimates of trade-related adjustment costs in Syria," *Journal of Policy Modeling* 32, pp. 843–864.

**Lloyd, P. J. and Zhang, X.-G. (2006),** *Thearmington model*. StaffWorking Paper 31921. Melbourne, Australia: Australian Productivity Commission.

**Lucke, B. (2001),** *Fiscal impact of trade liberalization: The case of Syria*, Proceedings of the 75th International Conference on Policy Modeling Brussels, Belgium.

**Mayer, W. (1984),** "Endogenous tariff formation," *American Economic Review* 74(5), pp. 970-985.

**Mitra, P. (1992),** "The coordinated reform of tariffs and indirect taxes," *World Bank Research Observer* 7(2), pp. 195–218.

**Romagnoli, A., and Mengoni, L. (2009),** "The challenge of economic integration in the MENA region: From GAFTA and EU-MFTA to small scale Arab unions," *Economic Change and Restructuring* 42(1–2), pp. 69–83.

**Pupongsak, S. (2009),** *The effect of trade liberalization on taxation and government revenue*, A thesis submitted to the University of Birmingham, Department of Economics.

**Tanzi, V., and Howell Z., (2001),** *Tax Policy for Developing Countries*, IMF, Washington D.C., 2001. Preface available at: http://www.imf.org/external/pubs/ft/issues/issues27/index.htm